\documentclass[acmlarge,screen]{acmart}

\AtBeginDocument{%
  \providecommand\BibTeX{{%
    \normalfont B\kern-0.5em{\scshape i\kern-0.25em b}\kern-0.8em\TeX}}}

\setcopyright{none}

\usepackage{xspace}

\usepackage[ruled,vlined,linesnumbered]{algorithm2e}
\usepackage{algorithmic}  
\usepackage{listings}
\usepackage{amsmath}
\usepackage{enumitem}
\usepackage{tabu}                      
\usepackage{booktabs}                  

\acmSubmissionID{0}

\begin{document}

\title[Reverse-Engineering Information Presentations]{Reverse-Engineering Information Presentations: \\ Recovering Hierarchical Grouping from Layouts of Visual Elements}


\author{Danqing Shi}
\affiliation{%
  \institution{Tongji University}
  \country{China}}
\email{sdq@tongji.edu.cn}

\author{Weiwei Cui}
\affiliation{%
  \institution{Microsoft Research Asia}
  \country{China}}
\email{weiwei.cui@microsoft.com}

\author{Danqing Huang}
\affiliation{%
  \institution{Microsoft Research Asia}
  \country{China}}
\email{danqing.huang@microsoft.com}

\author{Haidong Zhang}
\affiliation{%
  \institution{Microsoft Research Asia}
  \country{China}}
\email{haidong.zhang@microsoft.com}

\author{Nan Cao}
\affiliation{%
  \institution{Tongji University}
  \country{China}}
\email{nan.cao@gmail.com}

\renewcommand{\shortauthors}{Shi, et al.}


\begin{abstract}
Visual elements in an information presentation are often spatially and semantically grouped hierarchically for effective message delivery. Studying the hierarchical grouping information can help researchers and designers better explore layout structures and understand design demographics. However, recovering hierarchical grouping is challenging due to a large number of possibilities for compositing visual elements into a single-page design. This paper introduces an automatic approach that takes the layout of visual elements as input and returns the hierarchical grouping as output. To understand information presentations, we first contribute a dataset of 23,072 information presentations with diverse layouts to the community. Next, we propose our technique with a Transformer-based model to predict relatedness between visual elements and a bottom-up algorithm to produce the hierarchical grouping. Finally, we evaluate our technique through a technical experiment and a user study with 30 designers. The results show that the proposed technique is promising.
\end{abstract}




\maketitle
\section{Introduction}

Information presentations (e.g., presentation slides, business reports, and advertising posters) are widely used in daily life as an essential medium of visual communication. In an information presentation, the \textbf{layout} refers to the 2D spatial structure which is determined by the sizes and positions of the visual elements~\cite{lok2001survey}. To ensure effective message delivery, the layout is usually arranged by a set of design guidelines or principles. For example, Williams Robin provided practical design guidelines (e.g., proximity, alignment, repetition, and contrast) to suggest the layout when creating an information presentation~\cite{williams2015non}. More theoretically, the Gestalt laws provide the fundamental theory to explain the mechanisms of how human beings perceive and understand visual information~\cite{desolneux2004gestalt}. In a word, a well-designed layout structure enables the quick and easy perception of the information~\cite{wright1999psychology, rosenholtz2009intuitive}.

\begin{figure}[!t]
  \centering
  \includegraphics[width=\linewidth]{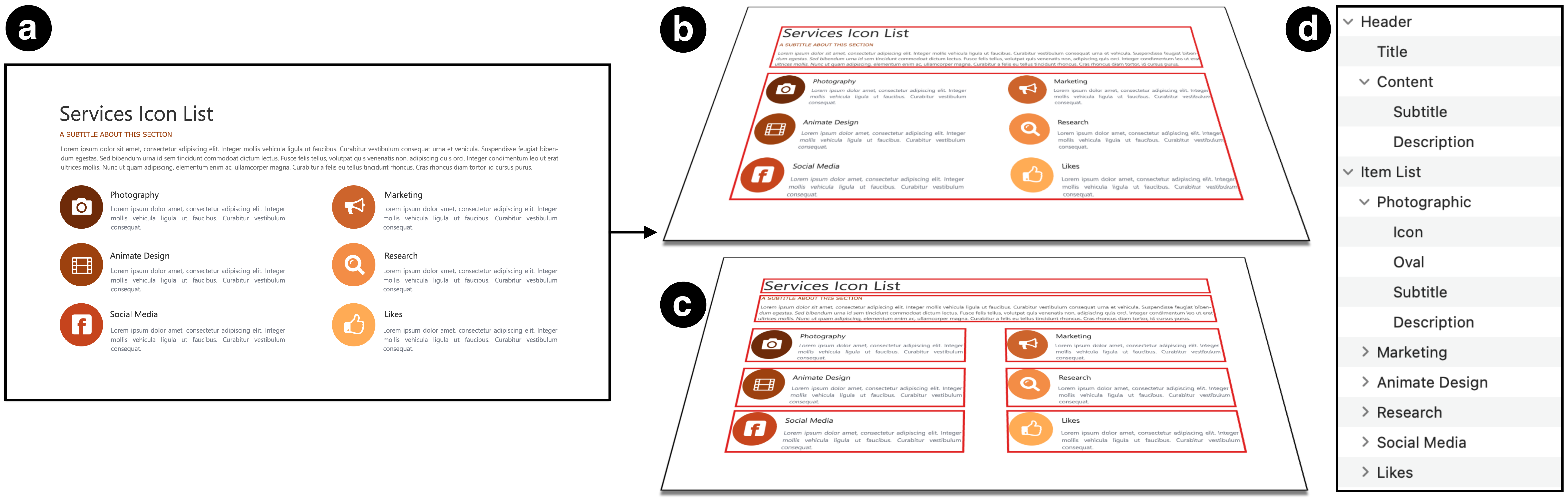}
  \caption{An example to show the grouping of information presented in two hierarchies (a) Original design; (b) Large grouping granularity; (c) Smaller groups; (d) The pane to view the hierarchical grouping in Powerpoint. High-level groups consist of low-level groups, which consist of visual elements. In each grouping hierarchy, visual elements in the same red box are in the same group.}
\label{fig:example}
\vspace{-1em}
\end{figure}

Hierarchical grouping is a popular approach to explaining the perceptual structure of the layout, which is commonly used in professional graphic design tools (see Fig.~\ref{fig:example} for an example). Here, a \textbf{group} is defined as a set of related visual elements that are associated with each other, and the \textbf{hierarchical grouping} is the collection of groups that are organized in a hierarchical structure. A proper hierarchical grouping can benefit many downstream use cases in visual design. A primary and direct use case is assisting users in designing information presentations easier: if the visual elements of an information presentation are grouped into a meaningful hierarchical structure, users can conveniently move/resize/rotate/transform/delete a group of related elements. It can also help to apply animation effects to the elements~\cite{wang2021animated}. Furthermore, other potential use cases that rely on understanding the grouping structure include design assessment~\cite{xu2014voyant}, visual search~\cite{hornof2001visual}, information extraction~\cite{poco2017reverse}, information seeking~\cite{hoque2019searching}, and design space exploration~\cite{lu2020exploring}.

Although useful for downstream applications, most information presentations in the wild do not explicitly annotate the hierarchical grouping. 
To address this problem, we aim to automatically recover hierarchical grouping from layouts of visual elements in information presentations. It is a challenging task for two major reasons. First, unlike the spatial relationships between elements in the layout such as direction and distance, which can be computed by using pre-defined rules~\cite{kong2006spatial}, the grouping structure is the implicit knowledge hidden in the layout. Second, there are many factors that may affect the grouping structure, which is hard for researchers to enumerate all the rules and principles. 
Several existing works have already paid attention to recovering grouping from arbitrary images~\cite{rosenholtz2009intuitive} and visualizations~\cite{wattenberg2003model}, but not from layouts of visual elements in information presentations. One of the most related works~\cite{koch2016computational} implements four Gestalt laws to produce grouping in web page design which performs well in their test cases. However, extensive human efforts are still required for researchers to determine better design rules to improve accuracy. Therefore, our emphasis in this paper is to incorporate a learning-based approach to a more intelligent grouping technique.

The large collection of information presentations available online serves as a great resource for our research. Therefore, we first constructed a large-scale dataset of information presentations. Specifically, as PowerPoint (denoted as PPT) is one of the most commonly used formats for information presentation in daily life, we have collected approximately 1,000 PPT files from online public websites, which finally gave us 23,072 well-designed information presentations. Each information presentation contains approximately 27 visual elements on average. By parsing visual elements from these slides, we extracted typical types of elements, such as shape, text, picture, and chart. To assist in understanding the layout, we also parsed all spatial features of each visual element.

Next, we propose an automatic approach to produce a hierarchical grouping of visual elements from information presentations. To address the problem by not enumerating all possible factors, we first incorporate a learning-based approach to model the relatedness between visual elements in information presentations. Specifically, we introduce a Transformer-based model, which can capture the contextual layout information to predict relatedness between visual elements by learning from human-labeled information presentations. In addition, we improve the original Transformer-based model by integrating the spatial-aware attention mechanism to involve the relative spatial information among elements. Second, our technique hierarchically groups all the elements from bottom to top according to the predicted relatedness. In detail, we define a graph model of the information presentation layout, in which nodes are visual elements and edges are the adjacent connections between elements. Then, we build a graph-based algorithm to group the visual elements hierarchically.

To evaluate the technique, we conduct a technical evaluation to compare the accuracy results of our models with a baseline method. We find that our technique can correctly predict pairwise relatedness with high accuracy. Furthermore, we conduct a user study with thirty designers to evaluate the final hierarchical grouping results. The results show that our technique has comparable quality with the slides manually grouped by designers. 

To summarize, we make the following three key contributions in this paper:

\begin{itemize}
    \item[1)] \textbf{Dataset.} We contribute a large-scale design dataset to the community. It includes 23,072 information presentations with 620,878 visual elements collected from the Internet. The dataset is freely available for academic purposes.
    \item[2)] \textbf{Technique.} We present an automatic approach for recovering hierarchical grouping in information presentations. It consists of two major modules: a Transformer-based model for predicting relatedness between visual elements and a bottom-up algorithm for grouping the visual elements hierarchically.
    \item[3)] \textbf{Evaluation.} We conduct a technical evaluation and a user study to evaluate our proposed technique. The results show that our technique can produce accurate hierarchical grouping results.
\end{itemize}

\section{Background and Related Work}

This section provides three categories of related work: graphic design layout, grouping of information, and data construction for graphic design.

\subsection{Graphic Design Layout}

To create effective and aesthetic information presentations, designers need to put significant effort into designing the layout~\cite{wright1999psychology}. Therefore, an extensive range of techniques has been proposed to assist layout design. 
Some studies in the community developed automatic layout algorithms that can place graphical and textual elements in proper positions based on certain design rules~\cite{lok2001survey}. For example, DesignScape~\cite{o2015designscape} is an interactive system that aids the design process by interactively generating layout suggestions, such as position, size, and alignment. SketchPlore~\cite{todi2016sketchplore} uses a real-time layout optimizer that automatically infers the designer’s task to search for both local design improvements and global layout alternatives. As the design of information presentation also commonly uses grid layouts to define spatial structures using grid lines~\cite{lupton2014thinking}, a more recent work GRIDS~\cite{dayama2020grids} uses a mixed-integer linear programming model for grid layout generation that optimizes the packing, alignment, grouping, and preferential positioning of elements. 
However, these rule-based methods typically aim to optimize certain objectives, which may ignore other aspects of the design. Recent advances in deep learning enable advanced techniques to generate graphic layouts by learning existing works. Zheng et al.~\cite{zheng2019content} proposed a deep generative model to synthesize layouts based on the semantic meaning of visual elements. Vinci~\cite{guo2021vinci} uses a sequence-to-sequence generative model to match the visual elements and layouts for generating an aesthetic advertising poster. LayoutGAN~\cite{li2020layoutgan} synthesizes layouts from different types of 2D visual elements by using a generative adversarial network (GAN) architecture. While these techniques are powerful in generating information presentation, they usually ignore the stage of understanding the layouts exactly.

Other works in the community focus on understanding the layout in the graphic design. For instance, Bylinskii et al.~\cite{bylinskii2017learning} presented a machine learning model that predicts the relative importance of different visual elements in graphic designs. Because inferring spatial relationships among elements plays a vital role in layout suggestion, Xu et al.~\cite{xu2014global} designed an interactive user interface for visualizing two types of spatial relationships: edge alignment and equal spacing. O’Donovan et al.~\cite{o2014learning} proposed an energy-based method to learn three hidden variables in single-page layouts, including alignment, the importance of each element, and grid-based segmentation.  These works are relevant to our work, which also extract the layout-related information from the design of information presentations. However, unlike these previous works, we aim to focus on hierarchical grouping, which is the implicit knowledge hidden in the layout of visual elements. In this work, we take advantage of a deep-learning-based approach to reverse-engineer information presentations, especially recovering the hierarchical grouping from layouts. 

\subsection{Grouping of Visual Information}

Although layouts might be vary specific to different document types, such as presentation slides, web pages, and advertising posters, the grouping of information is general to organize the content of information presentations. A proper grouping can improve the readability and highlight relatedness between the information~\cite{jansen1998graphical}. In practice, the \textit{proximity} guideline proposed by \cite{williams2015non} states ``Group related items together". Designers can move visual elements spatially close to make them a cohesive group rather than unrelated items. In other words, spatial closeness usually implies a high relatedness between the elements. In addition, the Gestalt principle states that human beings generally organize information using a top-down approach~\cite{wickens2015engineering, jansen1998graphical}. This principle suggests that a designer should organize visual information in a clear hierarchical grouping structure. 

Automatic grouping methods are extensively utilized in image processing. Ren et al.~\cite{ren2003learning} used the features derived from the
classical Gestalt cues to group pixels into regions. Wattenberg et al.~\cite{wattenberg2003model} utilized machine vision techniques for analyzing an image at multiple resolutions to describe the hierarchical segmentation of an image. Several bottom-up approaches were proposed for hierarchical grouping pixels in images or videos~\cite{arbelaez2014multiscale, pont2016multiscale, xu2016actor}. Recently, advanced deep learning methods have also been proposed to provide a range of end-to-end solutions for pixel-level grouping~\cite{gadde2016superpixel, jampani2018superpixel, li2020deep}. Nevertheless, these grouping techniques in image processing are all based on pixels. In contrast, grouping in information presentation focuses on the visual elements in a 2D view, which have more diverse spatial features.

There are several relevant grouping techniques for graphic design. Rosenholtz et al.~\cite{rosenholtz2009intuitive} utilized Gestalt laws to create a simple model for perceptual grouping for arbitrary images. Zhen et al.~\cite{xu2016identifying} established a model to merge web page content into semantic blocks by simulating human perception. Yu et al.~\cite{chen2003detecting} proposed a process to analyze the structure of a web page and organize the web elements into a two-level hierarchy. However, these web-oriented grouping techniques commonly require a formal HTML structure.
Recently in infographic design, InfoVIF~\cite{lu2020exploring} analyzed the presentation flow of static infographics and extracted the visual groups using Gestalt principles of proximity and similarity. InfoMotion~\cite{wang2021animated} extends InfoVIF to reconstruct visual groups using a clustering algorithm. However, these two recent techniques use carefully designed rules to pick elements into a visual group, which may ignore other design aspects. In addition, they both focus on the flow structure extraction but not the hierarchy in the grouping of information. 
One similar work introduced a computational method to group DOM elements in web pages~\cite{koch2016computational}. Nevertheless, they manually defined four objective functions that are hard to extend and improve. Another similar work by O’Donovan et al.~\cite{o2014learning} proposed a region-based hierarchical segmentation. The algorithm takes graphic layouts as input and then cuts the region into two sub-regions iteratively. This approach requires the design to follow a grid layout, and there should be non-overlapping visual elements. Compared to these works, our approach does not have the constraints for manually defined rules or strict requirements for the layout.

\subsection{Data Collection for Layout Design}

To support research on layout design in information presentations, a range of works construct datasets by collecting designs from public online websites. In the data visualization community, Chen et al.~\cite{chen2020composition} built a dataset containing 360 images of multiple visualizations collected from academic publications and made annotations of layouts. InfoVIF~\cite{lu2020exploring} provides a large dataset of 13,245 infographics in image format created to explore the visual information structure in infographics. Deka et al.~\cite{deka2017rico} presented Rico, the largest repository of mobile app designs by crowdsourcing, created to support data-driven applications such as UI layout generation. Webzeitgeist~\cite{kumar2013webzeitgeist} is another design repository of over 100,000 web pages with the statistical description of layouts. To support layout generation, Zheng et al.~\cite{zheng2019content} collected a corpus of 3,919 online magazine pages with rich annotations, including categories, semantic layouts, and textual summaries.
Compared with these studies, we aim to construct a large-scale layout dataset that contains rich spatial features to understand the layout structure. Therefore, we collected 23,072 well-designed information presentations from public websites. We parsed all the related spatial features of each visual element.
\section{Interviewing Experts}

To help us understand the hierarchical grouping used in the design, we conducted a preliminary interview with ten graphic designers (eight females, aged 23-34). These designers have 6.3 years of design experience on average (from 2 to 16 years) and have professional training in design schools. They are all familiar with designing information presentations. During the interview, we guided them to explain their design experience with questions under two primary topics: benefits from hierarchical grouping and criteria for hierarchical grouping. We summarize their feedback as follows:

\textbf{1) Benefits from hierarchical grouping}

As most professional graphic design software (i.e., Adobe Illustrator, Sketch) supports the hierarchical grouping, all the interviewees are familiar with this function. According to our interviewees, there are two reasons why they use hierarchical grouping in their information presentation files. The first major reason (mentioned by eight interviewees) is efficiently supporting layout design. For example, if a designer wants to align two groups of items, he or she can easily select one group at the proper level of hierarchy and then align it to another group. Otherwise, he or she has to operate each item in the group, which can absolutely reduce the efficiency. As another example, one expert in the interview said: ``\textit{I often need to create Web UI designs with various sizes to accommodate to different screen sizes, such as mobile, tablet, and laptop. A proper hierarchical grouping can save me lots of time to revise the layout}''. Second, four interviewees stated that hierarchical grouping can clearly segment a graphic design, assisting collaboration with other designers on the same document.
On the other hand, all the designers agreed that a design without a proper hierarchical grouping may cause trouble. Specifically, one designer commented that ``\textit{Without proper grouping, it is more possible to make mistakes when adjusting the layout and users have to undo the operation.}'' The designers also state that when reusing the design files in the wild, they have to manually adjust the hierarchical grouping of the visual elements before editing, which may burden their design process.

\textbf{2) Criteria for hierarchical grouping}

When discussing the best practice of grouping, designers mentioned several criteria. 
First, the items in a cohesive unit should have a high probability of being grouped together. For the example of Fig~\ref{fig:example}(a), the item in the list, which consists of an icon, a title, and a description to convey a single message, should be grouped together because these elements form a highly cohesive unit. 
Second, designers agreed among themselves on the significance of spatial distance when grouping elements. Elements relating to each other should be grouped close together. Designers also admitted that they only considered grouping elements in the neighborhood in most situations. 
Finally, the grouping was suggested to follow a visual hierarchy, which plays a key role in the planning of the layout to help viewers navigate through the information presentation more easily.


\textbf{Design requirements}.
Our goal is to propose an automatic approach to recovering hierarchical grouping from information presentations.
To achieve the goal, we summarize two design requirements based on the feedback gathered from the interviews with designers.

{\bf R1} {\bf Capturing relatedness.} 
The first requirement is to capture relatedness between visual elements, which can be affected by many factors, including types and spatial features. 
The technique should be intelligent enough to predict relatedness by considering as much information as possible.

{\bf R2} {\bf Constructing a hierarchical structure.}
The grouping result should be a hierarchical tree structure according to the visual hierarchy. A parent group can contain multiple sub-groups. However, a sub-group cannot be across multiple parent groups. In this way, the exact group can be easily accessed from its parent group.

To fulfill these requirements, the design of our technique consists of two major modules (1) a deep learning model for predicting pairwise relatedness (Section~\ref{sec:model}) and (2) a graph-based algorithm for grouping (Section~\ref{sec:algorithm}). In the first module, the deep learning model can understand the relatedness between visual elements by learning from labeled training data ({\bf R1}). In the second module, we model the layout as a graph model to represent the adjacent connection. Then, we propose a graph-based algorithm to group the visual elements from bottom to top hierarchically ({\bf R2}). 
\section{Collecting Information Presentations}

To understand information presentations, we construct a large-scale dataset of information presentations with diverse layouts from the Internet. First, we describe the procedure to collect the data. Then, we analyze the statistical results and extract the spatial features.

\subsection{Collection Procedure}

We chose PowerPoint files as the target file format to construct our dataset because they are the most commonly used information presentations. Each slide in the PPT file is an information presentation that consists of a set of graphical and textual elements. Furthermore, PPT files contain the actual spatial information for each visual element, which can be parsed by an open-source tool~\footnote{https://pypi.org/project/python-pptx}. Therefore, we can obtain accurate positions of visual elements from information presentations without using any computer-vision methods such as InfoVIF~\cite{lu2020exploring}.

To ensure the design quality and diversity, we collected PPT files through the two key steps: (1) searching a large number of candidates in the wild and (2) selecting proper slides to include in our dataset.

In the first step, we searched for public websites such as SlideShare~\footnote{https://www.slideshare.net} that are used for sharing slides. We also searched related keywords such as ``slide template'', ``marketing slides'' and ``presentation design'' in popular search engines to collect open-source PPT files. We include as many slides as possible in this step. At the end of this step, we downloaded approximately 1,000 PPTs in total, including more than 40,000 information presentations.

In the second step, we selected infographics-related information presentations from these candidates to construct our dataset. To ensure that the layouts of information presentations are diverse and qualified, we define the following four criteria to filter the slides. 

\begin{figure}[!t]
  \centering
  \includegraphics[width=\linewidth]{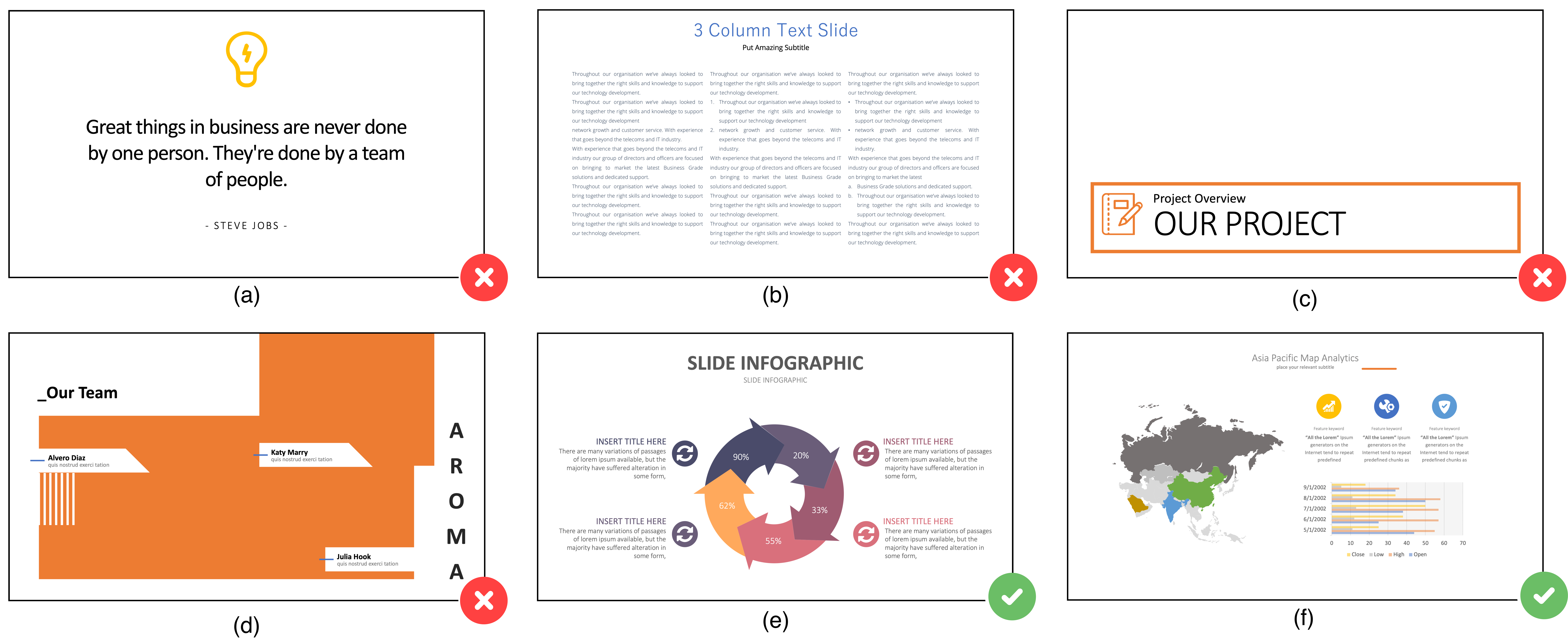}
  \caption{An illustration of four negative cases that we removed from the dataset: (a) includes only three visual elements (one icon and two textual objects); (b) has no graphical elements; (c) just contains one visual group; (d) has a disorder layout. (e) and (f) are two positive cases in the dataset.}
\label{fig:cases}
\vspace{-1em}
\end{figure}

{\bf C1} To ensure that visual elements of an information presentation are worth grouping, we ask that slides should have more than three visual elements. As the counterexample is shown in Fig~\ref{fig:cases}(a), it only includes an icon shape and two textual objects. Too few elements are not valuable to be grouped.

{\bf C2} To ensure the diversity of layouts, we ask that slides should contain at least two visual groups. As we can see in the counterexample in Fig~\ref{fig:cases}(b), which is a title page, the four elements (a rectangle, an icon shape, and two textual objects) form a coherent group to show a single message.

{\bf C3} To ensure an information presentation is visually attractive and informative, we ask that slides should include both graphical and textual elements. The counterexample in Fig~\ref{fig:cases}(c) presents a slide with only text information.

{\bf C4} The elements in the slide should be arranged in a well-organized structure. It also means the elements should form an intuitively regular pattern~\cite{lu2020exploring}. In Fig~\ref{fig:cases}(d), a counterexample shows a layout whose elements are disordered to some extent.

We wrote a simple program to filter out invalid slides (C1 and C3). As C2 and C4 are related to subjective judgment, human effort is involved in assessing whether a slide is qualified. Two authors of this paper manually filtered these slides. Both of them are experts in designing information presentations. One has studied in a design school for three years, and the other has worked on information presentation for over five years. Each slide takes one author approximately 5 seconds to decide whether to keep it or remove it. They kept the cases they were unsure about and met every three days to discuss these unsure cases and achieve consensus. The whole process took seven days (about six hours per day) to look through all the slides. Fig~\ref{fig:distribution}(e) and Fig~\ref{fig:distribution}(f) are two typical information presentations included in our dataset. They have diverse visual elements and clear layout structures.

\subsection{Results}

\begin{figure}[!tbh]
  \centering
  \includegraphics[width=\linewidth]{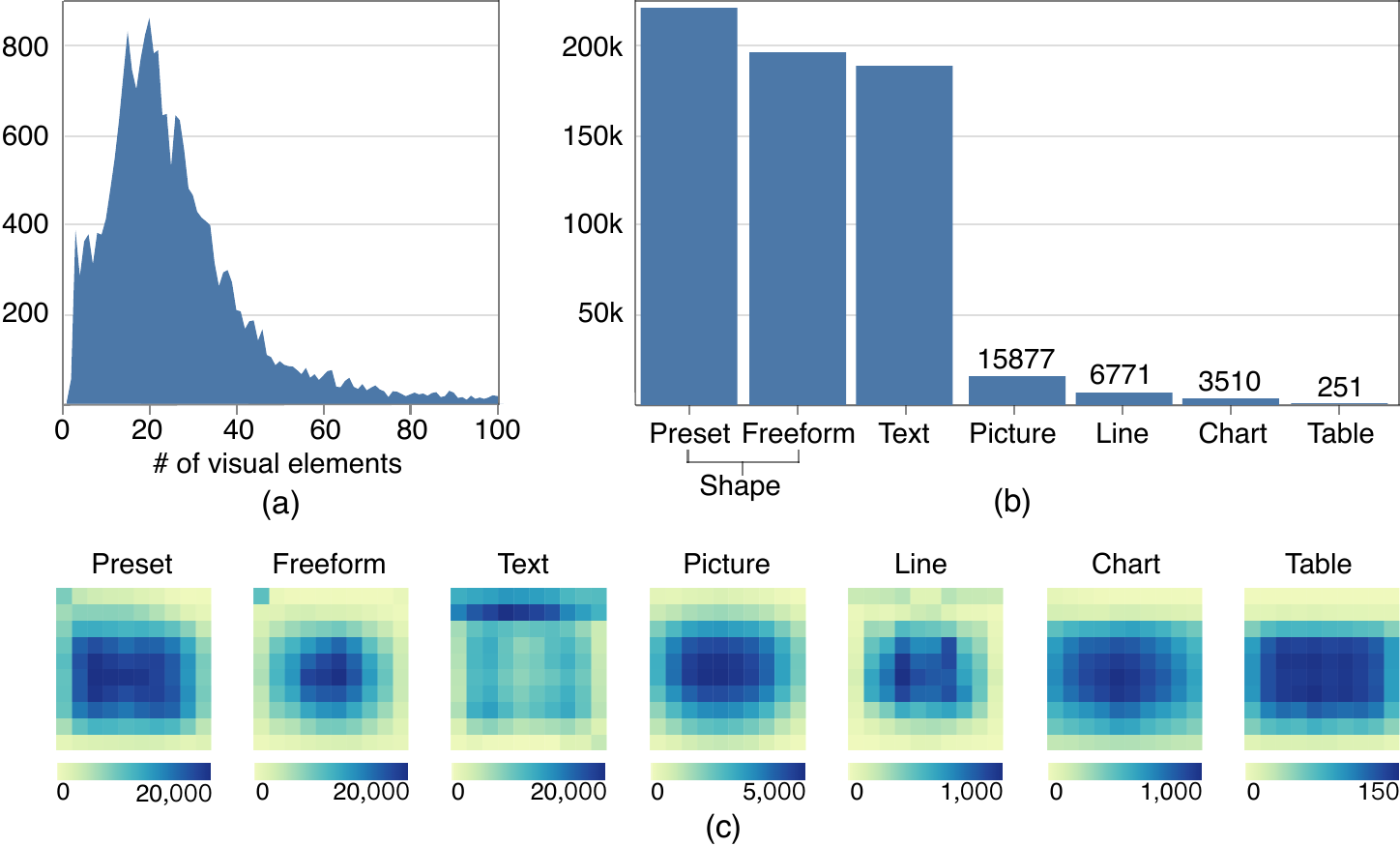}
  \caption{Summary statistics. (a) The distribution of element count per slide; (b) The distribution of types in the dataset; (c) Spatial distribution of the seven element types on the normalized space. The color of a region indicates the count of the element type appearing in the corresponding region.}
\label{fig:distribution}
\vspace{-1em}
\end{figure}

We finally collected a total of 23,072 slides with 620,878 visual elements. The average number in each slide is 26.9. Fig~\ref{fig:distribution}(a) presents the distribution of the number of elements in our data collection. The majority of the slides (83.7\%) contain less than 40 elements. More than half of the slides have 10 to 30 visual elements (57.3\%). 

We encode the layout structure of each slide with different \textit{type}s of visual elements. In total, there are seven major \textit{type}s we extracted from our collection: \textit{preset geometry}, \textit{freeform shape}, \textit{text}, \textit{picture}, \textit{line}, \textit{chart}, and \textit{table}. The distribution of element types is displayed in Fig~\ref{fig:distribution}(b). Let us look at the meaning of all the element types. A \textit{preset geometry} (216,300) is a typical shape provided by general design tools, such as a rectangle, a circle or a star. In detail, the most commonly used \textit{preset geometry} is rectangle (119,818), and the second is oval (56,560). A \textit{freeform shape} (192,673) is a user-defined shape that is composed of multiple paths. These kinds of elements are usually icons or embellishments in the design, which might have a related semantic meaning for the design. A \textit{text} (185,259) is a text box that enables users to input text information. It is the most important component that is responsible for conveying messages in information presentations. The other four types are not the majority of the visual elements: \textit{picture} (15,887), \textit{line} (6,771), \textit{chart} (3,510) and \textit{table} (251). Fig~\ref{fig:distribution}(c) illustrates the spatial distributions of these seven element types on the normalized space. This demonstrates the diversity of our data collection in terms of layouts of visual elements. From these statistical results, we can see that the information presentations in our data collection have rich visual elements and diverse layout structures.

Then, we extracted related spatial features from each visual element. Specifically, we recorded the element type and the spatial features, covering the z-index, position, size, and rotation. For the textual elements, we also extracted the alignment of the text. The reason to pick these features is that all of them are highly relevant to the spatial structure. We describe these spatial features as follows:

\begin{itemize}
    \item[1)] \textit{Z-index} indicates the order of elements in the z-axis. An element with a higher z-index is on top of another element with a lower z-index. We represent the z-index using a natural number. 
    \item[2)] \textit{Position} indicates the position of the element. We use the bounding box $\left(x_1, y_1, x_2, y_2\right)$ to represent the position of the element, where $\left(x_1, y_1\right)$ denotes the top left of the bounding box and $\left(x_2, y_2\right)$ denotes the bottom right. We adopt the normalized parameters by scaling $\left(x_1, y_1, x_2, y_2\right)$ to the range of [0, 1]. 
    \item[3)] \textit{Size} $\left(w, h\right)$ is the normalized width and height of a visual element. Although this feature can be computed by the position feature, we explicitly represent this feature to help the computer understand this relationship more easily.
    \item[4)] \textit{Rotation} indicates the degrees of clockwise rotation of the element ranging from 0 to 360. 
    \item[5)] \textit{Alignment} is a text-specific feature as a text can be aligned in the vertical and horizontal directions. It includes two attributes for vertical alignment (top, middle, bottom, and mixed aligned) and horizontal alignment (left, right, center, and mixed aligned). 
\end{itemize}

To protect data privacy, we removed all the content-related information, such as text and images. The final dataset only keeps the features of each visual element mentioned before, which is enough for the study of layout structure. We open-sourced the dataset~\footnote{https://github.com/sdq/reip}. More discussion about the usage of the dataset in Section~\ref{sec:usage}.
\section{Predicting Pairwise Relatedness}
\label{sec:model}

Before we group visual elements in an information presentation, we first consider the relatedness between elements, which also implies the probability of grouping two elements together ({\bf R1}). This section introduces a deep learning model to solve this problem. First, the basic model architecture and the working mechanism is presented. Next, we introduce an improvement to enhance the performance. Finally, the training process and model implementation is briefly described.

\subsection{Model Architecture}

To predict the relatedness between any two visual elements in an information presentation, we propose a Transformer-based~\cite{vaswani2017attention} model for pairwise prediction. 
There are two main reasons for doing so. First, as an information presentation could be formalized as a design sequence~\cite{guo2021vinci}, it is reasonable to adopt a sequence model in our task. Compared with traditional sequence models such as recurrent neural networks, the self-attention mechanism~\cite{vaswani2017attention} in Transformer models can capture long-range contextual information in the whole layout design. Specifically, understanding the spatial information of all visual elements in the design as global context is helpful when predicting the relatedness between two visual elements. Second, the language model on the Transformer architecture has been widely adopted. It would be easy to adapt the pre-training to have a promising improvement with the growth of data collection.

\begin{figure}[!t]
  \centering
  \includegraphics[width=\linewidth]{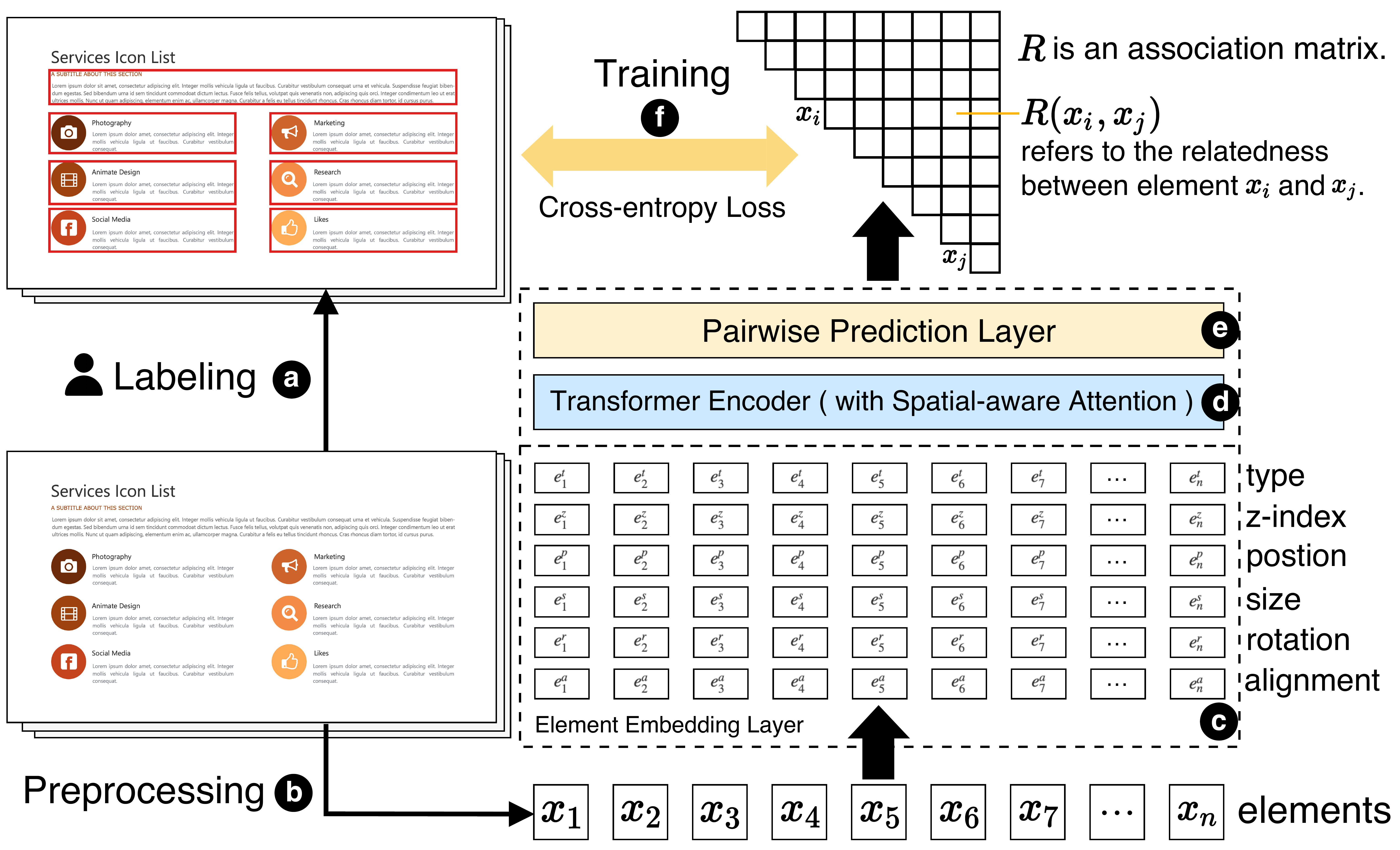}
  \caption{The model architecture and mechanism for predicting relatedness between visual elements}
\label{fig:model}
\vspace{-1em}
\end{figure}

The model architecture is demonstrated in Fig~\ref{fig:model}. It relies on supervised training based on human-labeled information presentations (Fig~\ref{fig:model}(a)). The input of the model is a series of visual elements parsed from the information presentation (Fig~\ref{fig:model}(b)), and the output is an association matrix of predicted relatedness between elements. The core model is based on the Transformer backbone, which is similar to the latest work for learning layout representation~\cite{xie2021canvasemb}. It consists of three components: (1) an embedding layer to represent visual elements with heterogeneous features (Fig~\ref{fig:model}(c)), (2) a Transformer encoder for contextualized representation (Fig~\ref{fig:model}(d)), and (3) a prediction layer to predict pairwise relatedness (Fig~\ref{fig:model}(e)). The model is trained to minimize the cross-entropy loss between the output and the ground truth (Fig~\ref{fig:model}(f)).

\subsection{Working Mechanism}


\textbf{Pre-processing}. 
Consider the example layout design in Fig~\ref{fig:model}(a). The pre-processing module (Fig~\ref{fig:model}(b)) first parses the design file to extract all visual elements in the design and then orders it into a sequence of visual elements according to the z-index. We represent the design sequence as: $\left[x_1, x_2, ..., x_n\right]$, where $x_i$ indicates the $i$-th element in the sequence. According to the statistics in Section 4.2, we set the maximum sequence length to 128, as most of the slides in our dataset contain 128 elements or fewer. For efficient batch training, we fill the design sequence to this maximum length with special padding tokens.

\textbf{Embedding}. 
To abstract visual elements into a data format that is understandable to the Transformer model, we convert each visual element $x_i$ into a set of feature embedding vectors. For the numerical features, including the \textit{z-index}, \textit{position}, \textit{size}, and \textit{rotation}, we embed them using the positional encoding method~\cite{vaswani2017attention}. For the remaining two categorical features \textit{type} and \textit{alignment}, we embed them using a learnable embedding table. The final representation of a visual element is a single embedding vector $e_i$ by concatenating all the feature embedding vectors as follows:

\begin{equation}
e_i = \Theta ( e^t_i \oplus e^z_i \oplus e^p_i \oplus e^s_i \oplus e^r_i \oplus e^a_i )
\end{equation}
where $e^t_i$ to $e^a_i$ denote the feature embedding vectors for \textit{type}, \textit{z-index}, \textit{position}, \textit{size}, \textit{rotation}, and \textit{alignment}, respectively. $\Theta$ is a non-linear transform function. 
In this way, the output of the embedding layer is a sequence of visual element embeddings: $\left[e_1, e_2, ..., e_n\right]$, which will serve as the input to the Transformer encoder.

\textbf{Encoding}.
The Transformer encoder (Fig~\ref{fig:model}(d)) updates the element embeddings into contextualized vectors $[h_1, h_2, ..., h_n]$. According to the query-key-value attention mechanism in Transformer, the model will involve three learnable weight matrices: query weight matrix $\mathbf{W}_Q$, key weight matrix $\mathbf{W}_K$, and value weight matrix $\mathbf{W}_V$. The self-attention mechanism computes the attention score between the query of element embedding $e_i$ and the key of element embedding $e_j$ as follows:

\begin{equation}
\alpha_{i j}=\frac{1}{\sqrt{d_{\text {k}}}}\left(\mathbf{e}_{i} \mathbf{W}_{Q}\right)\left(\mathbf{e}_{j} \mathbf{W}_{K}\right)^{\top}
\end{equation}
where $\sqrt{d_{\text {k}}}$ is the scaling factor to stabilize gradients during training. The output of the Transformer encoder is the weighted average of the value vectors of each element embedding ${e}_{j}$ with normalized attention scores:

\begin{equation}
\mathbf{h}_{i}=\sum_{j} \frac{\exp \left(\alpha_{i j}\right)}{\sum_{k} \exp \left(\alpha_{i k}\right)} \mathbf{e}_{j} \mathbf{W}_{V}
\end{equation}

\textbf{Predicting.}
In the prediction layer (Fig~\ref{fig:model}(e)), we predict the pairwise relatedness between every two elements. We use an association matrix $\mathbf{R}$ with size $(n \times n)$ to denote the prediction result. In the matrix, each item $\mathbf{R}\left(x_i, x_j\right)$ represents the relatedness between element $x_i$ and $x_j$ randing from 0 to 1, where 1 means the model is confident that the two elements should be assigned in the same group and 0 means there is no possibility to group them. The predicted relatedness $\mathbf{R}\left(x_i, x_j\right)$ is the average value of two related attention scores $r_{ij}$ and $r_{ji}$:

\begin{equation}
r_{ij} = \frac{\exp{\mathbf{W}_Q^{r}\textbf{h}_i \cdot \mathbf{W}_M^r\textbf{h}_j}}{\sum_{j}\exp{\mathbf{W}_Q^r\textbf{h}_i \cdot \mathbf{W}_M^r\textbf{h}_{j}}}
\end{equation}
where $\mathbf{W}_Q^{r}$ and $\mathbf{W}_M^{r}$ are two learnable matrices for query and memory according to the query-memory attention mechanism~\cite{wang2017gated}. The $r_{ij}$ denotes the attention score between the query of $\mathbf{h}_i$ and the memory of $\mathbf{h}_j$; \textit{vice versa} for $r_{ji}$.

\subsection{Improvement}

The spatial relations between elements, including the relative \textit{z-index} and relative \textit{position} in both the \textit{x}-axis and \textit{y}-axis, are informative for measuring the relatedness between elements.
However, the basic attention mechanism in the current Transformer encoder only considers the absolute position of visual elements, while the relative information is not explicitly modeled.

To better utilize this information, we improve the model by incorporating the spatial-aware attention mechanism\cite{shaw2018self, xu2020layoutlmv2}. Specifically, we compute a new attention score $\alpha_{i j}^{\prime}$ by adding three learnable biases (relative \textit{position} biases $\mathbf{b}^{\mathrm{X}}$ and $\mathbf{b}^{ \mathrm{Y}}$, relative \textit{z-index} bias $\mathbf{b}^{\mathrm{Z}}$) to the original score as follows:

\begin{equation}
\alpha_{i j}^{\prime}=\alpha_{i j} +\mathbf{b}^{
\mathrm{X}}+\mathbf{b}^{\mathrm{Y}} + \mathbf{b}^{ \mathrm{Z}}
\end{equation}

We have empirically verified that adding this spatial-aware attention mechanism can increase the accuracy of the model, which our technique will use by default. The evaluation of the improvement is presented in Section~\ref{sec:techevaluation}.

\subsection{Training} 
To train the prediction model, we first label the relatedness between elements as ground truth. The labeling process is simple: PowerPoint is used to group the related items (Fig~\ref{fig:model}(a)). Each pair of elements is labeled as a binary value (one if they are in the same group or zero if not). It is worth noting that this labeling process only needs a single grouping hierarchy, so that general users without a design background can easily label the slides based on their perception and cognition. In addition, we expand the size of labeled training data by crowdsourcing as the task is easy to perform. The training objective is to minimize the cross-entropy loss between the association matrix from the model and the ground truth matrix (Fig~\ref{fig:model}(f)). 

\subsection{Implementation Details}

The model was implemented in PyTorch~\cite{paszke2019pytorch}. All the trainable parameters are randomly initialized and updated via the Adam optimizer~\cite{kingma2014adam}, with a learning rate of 0.0001. The training process enables a 0.3 dropout rate to avoid over-fitting. The maximum length of the input design sequence is 128 elements. The batch size is set to 16. The model was trained on an Nvidia Tesla-V100 (16GB) GPU. 
\section{Grouping Visual Elements}
\label{sec:algorithm}

This section introduces a bottom-up approach to recover the hierarchical grouping from layouts of visual elements. We first define a graph model of the layout based on the connections of adjacent elements (\textbf{R2}) and then propose a graph-based algorithm to group the visual elements hierarchically (\textbf{R3}).

\subsection{Graph Modeling}

\begin{figure}[!h]
  \centering
  \includegraphics[width=\linewidth]{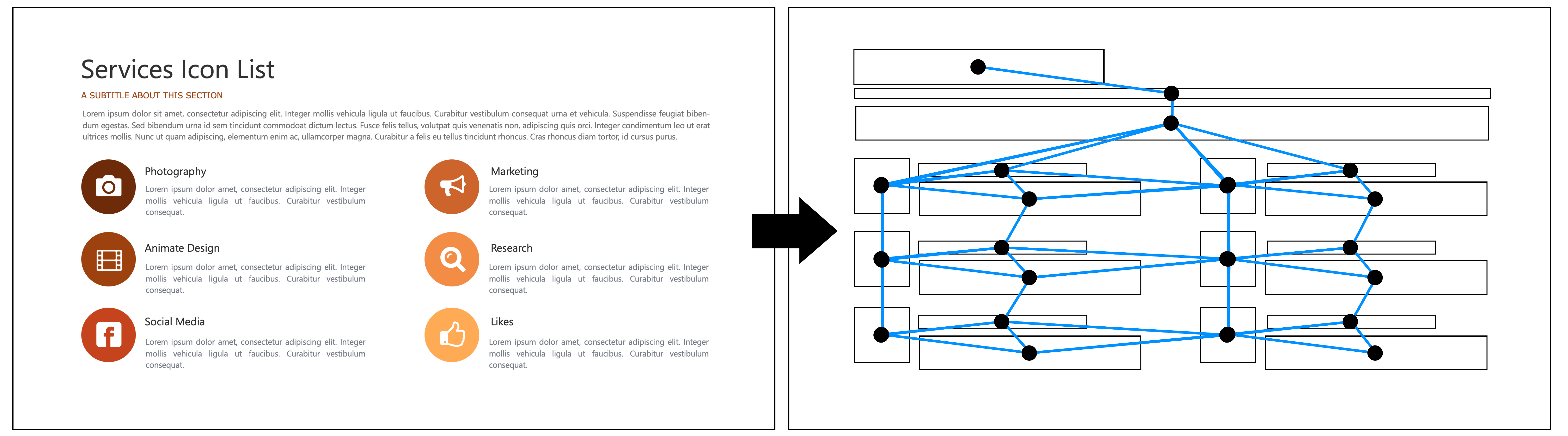}
  \caption{This is an illustration of the graph modeling for an information presentation, where nodes are visual elements (denoted as black circles), and edges are connections between two adjacent elements (denoted as blue lines).}
\label{fig:graph}
\vspace{-1em}
\end{figure}

According to the expert interviews, designers generally follow the principle of proximity to organize the layout and place related elements adjacently. Therefore, we only consider grouping elements in the neighborhood in our implementation. Thus, we can explicitly form a graph model to represent the layout using the spatial features of the visual elements  (Fig.~\ref{fig:graph}). Formally, we define the graph model $G$ of the information presentation, where nodes are visual elements $[x_1, x_2, ..., x_n]$ and edges are connections between the neighborhood elements, where we denote the edge that connects elements $x_i$ and $x_j$ as $(x_i, x_j)$. The edge includes a weight of proximity that can be computed by the spatial distance between two elements (denoted as $d(x_i,x_j)$). We use the largest distance among three perspectives: Euclidean distance, horizontal distance, and vertical distance~\cite{lu2020exploring}. 

\subsection{Hierarchical Grouping Algorithm}

Next, we use the bottom-up approach to group visual elements hierarchically. Our hierarchical grouping algorithm is inspired by the previous graph-based algorithm for image segmentation~\cite{grundmann2010efficient, felzenszwalb2004efficient}. Their objective is to group pixels or regions that have a similar visual appearance. Instead, our objective is to group visual elements relatedly and spatially. The basic idea of the algorithm is merging small groups into a larger group iteratively by considering relatedness and spatial distances between visual elements in different groups. The algorithm works as follows (Algorithm~\ref{alg:grouping}):

\begin{figure}[htb]
  \centering
  \begin{minipage}{0.6\linewidth}
    \begin{algorithm}[H]
      \label{alg:grouping}
    \SetAlgoLined
    \SetKwInOut{Input}{Input}
    \SetKwInOut{Output}{Output}
    \SetKw{KwBy}{by}
    \Input{graph model $G$, association matrix $\mathbf{R}$}
    \Output{hierarchies of grouping $\mathcal{H} = [H_0, H_1, ...]$}
    
    $H_0 \leftarrow [x_1, x_2, ..., x_n] \in G$; $\mathcal{H} \leftarrow [H_0]$;\\
    $edges \leftarrow {\rm sortByWeight}(edges \in G) $; \\
    $\tau \leftarrow \tau_{initial}$; $T \leftarrow T_{initial}$; \\
    \While{$|H_q| > 1$}{ 
        $q \leftarrow q+1$; $H_q \leftarrow H_{q-1}$; \\
        \For{$ (x_i, x_j) \in edges $}{
            \If{$x_i$ and $x_j$ in two different groups $\mathcal{G}_a$ and $\mathcal{G}_b$ }
            {
                $C_1 \leftarrow \mathbf{R}(x_i, x_j) \geqslant T $ \\
                $C_2 \leftarrow d(x_i, x_j) \leqslant {\rm MID}(\mathcal{G}_a, \mathcal{G}_b, \tau) $ ; \\
                \If{$C_1 \land C_2$}{
                    $H_q \leftarrow merge(\mathcal{G}_a, \mathcal{G}_b)$
                }
            }
        }
        \If{$|H_q| < |H_{q-1}|$}
        {
            $\mathcal{H} \leftarrow H_q$; \\
        }
        $T \leftarrow \alpha \times T $; $\tau \leftarrow \beta \times \tau$; \\ 
    }
    \Return $\mathcal{H}$\;
    \caption{Hierarchical Grouping Algorithm}
    \end{algorithm}
  \end{minipage}
\end{figure}

The input of the algorithm is the graph model with visual elements $[x_1, x_2, ..., x_n]$ as nodes and all connections as edges $[(x_1, x_2), ..., (x_i, x_j), ...]$. The outputs are levels of hierarchy $\mathcal{H} = [H_1, H_2, ..., H_l]$ from bottom to top, where $H_i$ contains the groups at the $i$th level of hierarchy.
We start with the initial grouping $H_0$, where each visual element is assigned to an individual group (\textit{line 1}). Next, we sort edges by non-decreasing spatial distance (\textit{line 2}) and initialize two parameters $\tau$ and $T$. $\tau$ is the threshold parameter that controls the degree to which the distance between two groups must be greater than their internal distances in order for there to be evidence of a boundary between them~\cite{felzenszwalb2004efficient}. $T$ is the threshold parameter for the relatedness (\textit{line 3}). These two parameters influence the granularity of the grouping. A larger $\tau$ and a smaller $T$ lead to a higher level of hierarchy with larger groups. 
After the initialization, the algorithm begins to merge small groups into a larger one iteratively until all the elements are in the one group (\textit{line 4-18}).
In each iteration, we enumerate all edges by order (\textit{line 6}). If two elements in an edge (e.g. $x_i$ and $x_j$) belong to two different groups (\textit{line 7}), we will consider whether to merge these two groups based on following two conditions:

\textbf{Relatedness.} The first condition is whether the predicted relatedness $R(x_i, x_j)$ between $x_i$ and $x_j$ is larger than the threshold $T$ (\textit{line 9}). $R(x_i, x_j)$ can be accessed from the association matrix $R$ from our model.

\textbf{Spatial distance.} The second condition is whether the spatial distance between $x_i$ and $x_j$ is smaller than the minimal internal distance of both groups (\textit{line 8}). The minimal internal distance ${\rm MID}$ is defined as:
\begin{equation}
\operatorname{MID} =\min \left(\operatorname{I}\left(\mathcal{G}_a\right)+\frac{\tau}{\operatorname{A}(\mathcal{G}_a)}, \operatorname{I}\left(\mathcal{G}_b\right)+\frac{\tau}{\operatorname{A}(\mathcal{G}_b)}\right)
\end{equation}
where $\operatorname{A}(\mathcal{G})$ computes the area size of the group, and $\operatorname{I}(\mathcal{G})$ computes the largest distance in the minimum spanning tree $\operatorname{MST}$ of the group $\mathcal{G}$: 
\begin{equation}
\operatorname{I}(\mathcal{G}) = \max _{(x_i, x_j) \in \operatorname{MST}(\mathcal{G})} d(x_i, x_j)
\end{equation}

If both conditions are true (\textit{line 10}), the two groups will be merged into a larger group (\textit{line 11}).
After each iteration, if the number of groups decreases, it means the algorithm constructs a higher-level grouping $H_q$ given the previous one $H_{q-1}$  (\textit{lines 15 - 17}). We will update the constant parameters by decreasing $T$ and increasing $\tau$ before starting the next iteration (\textit{line 18}). The constant parameters $\alpha$ and $\beta$ can be manually tuned to adapt to different types of information presentations. In our implementation in PowerPoint files, we set $\alpha = 0.9$ and $\beta = 1.1$.
Finally, the algorithm outputs all levels of hierarchy $\mathcal{H}$ (\textit{line 19}) that allow selection of the desired grouping at any desired granularity (\textbf{R3}).
\section{Evaluation}

In this section, we evaluate our technique, applied to hierarchical grouping of information presentations, through a technical experiment and a user study.

\subsection{Technical Experiment}
\label{sec:techevaluation}

We first present quantitative assessments of the prediction model. In the experiment, we manually labeled 1,008 slides from our data collection. 
Although it only includes a small portion of the data collection, it can create enough pairs of visual elements for training. The labeled slides include 27,151 visual elements, which comprise 519,415 element pairs in total. 
We divide the labeled data as 80/20 for training and testing. 


Given a pair of visual elements in an information presentation, our approach can predict the relatedness. If the relatedness score is higher than a threshold $t$ ($t=0.5$), our approach will predict that they are probably in the same group.
We use $accuracy$ as the evaluation metric because we have the human-labeled group truth. Formally, $accuracy$ is equal to $M/N$, where $M$ is the number of the correct predicted pairs, and $N$ is the total number of pairs in the testing data.

We compared the performances of the following three machine learning models on the testing dataset. 
Note that we did not involve other layout-based machine learning models here because most of the works focus on layout generation~\cite{li2020layoutgan, arroyo2021variational}, which is not applicable in the grouping task.

\begin{itemize}
    \item[1)] \textit{Neural Network Baseline.} We implemented a three-layer neural network as the baseline, which uses the same input feature embedding vectors of the visual elements and outputs the results of pairwise classification.
    \item[2)] \textit{Our Transformer Model without Spatial-aware Attention.} We implemented a basic version of our model without the spatial-aware attention mechanism. It shows the power of the original Transformer architecture.
    \item[3)] \textit{Our Transformer Model with Spatial-aware Attention.} We implemented the final version of our model by adding the spatial-aware attention mechanism, which can leverage the relative position information.
\end{itemize}

We summarize the performances of these three models in Table~\ref{tab:performance}. The neural network baseline approach achieves 77.49\% accuracy, while our approaches outperform it significantly by 8.23\% and 8.83\%, respectively. The result is expected because our models leverage the Transformer architecture to capture the contextual information from the whole layout structure. Next, let us look at the performance difference between our model with and without the spatial-aware attention mechanism. The accuracy of our final model (86.32\%) is slightly higher than that without spatial-aware attention (85.72\%), which indicates that using relative position information can improve the relatedness prediction in the 2D layout to some extent. The experiment shows the ability of our prediction model to capture the relatedness between visual elements in information presentations. 

\begin{table}[!t]
\fontsize{10}{10}\selectfont
\caption{Accuracy of the models for predicting relatedness between visual elements}
\begin{tabular}{p{5.8cm}p{1.8cm}}
\toprule
\textbf{Model} & \textbf{Accuracy} \\ \midrule
Neural Network Baseline & 77.49\% \\ 
Ours w/o Spatial-aware Attention & 85.72\% \\
Ours with Spatial-aware Attention & 86.32\% \\ 
\bottomrule
\end{tabular}
\label{tab:performance}
\vspace{-1em}
\end{table}

\subsection{User Study}

We conducted a user study to evaluate the quality of the generated hierarchies.
Although there are several grouping techniques for graphic design~\cite{lu2020exploring,wang2021animated}, they focus on structure extraction but not hierarchical grouping. Other related hierarchical grouping techniques~\cite{o2014learning,koch2016computational} require the design of information presentations following specific rules, such as using a grid layout or no overlaps among visual elements, which limit the scope of the technique used. 
Therefore, we directly compared the results with the baseline from human designers.   
The experiment consists of two major phases: (1) Ten designers manually grouped visual elements hierarchically as the baseline. (2) The other 20 designers rated the results from the baseline and our approach.
We hypothesized that our technique could produce a comparable result with the human designers.


\textbf{Study Data.} To prepare the study data, we first randomly sampled 50 slides from our dataset. We invited ten designers from our preliminary interview to manually group visual elements hierarchically in the slides. Before they started, we briefly introduced the grouping task and the basic usage of PowerPoint. During the task, each designer was given a PowerPoint file with ten different slides. They were asked to hierarchically group the elements in each slide. We record the completion time for each designer. Finally, each designer completed ten hierarchical grouping results, which gave us a total number of 100 labeled slides (one slide had two labeled versions from two random designers, respectively). An interesting finding is that designers may have different grouping solutions for the same information presentation.
We also used our technique to automatically recover hierarchical grouping from these slides. To compare the automatic results with the manually labeled results, we constructed 100 comparison pairs. Each pair includes one result from a designer and one result from our technique for the same slide. We ensure that the generated grouping result has the same number of hierarchies for a fair comparison by deleting excess hierarchies. 

\textbf{Participants.} We recruited 20 participants (14 females, aged from 22 to 31) from a design school. All the participants were familiar with design software such as Adobe Illustrator and Microsoft PowerPoint. They all confirmed having experience in making information presentations using Microsoft PowerPoint.

\begin{figure}[!t]
  \centering
  \includegraphics[width=\linewidth]{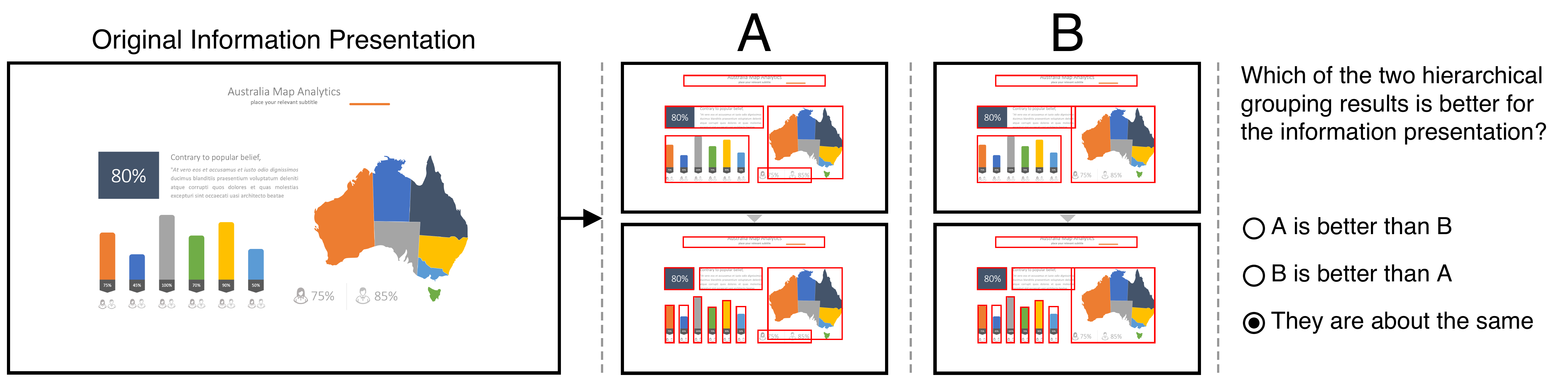}
  \caption{A snapshot of a side-by-side comparison pair used in the evaluation. (a) is manually grouped by a designer, and (b) is automatically recovered from our technique.}
\label{fig:studydemo}
\vspace{-1em}
\end{figure}

\textbf{Task and Procedure.} The experiment began with a brief introduction. Each participant was asked to complete 25 comparison tasks. In each comparison task, the participants were given two hierarchical grouping results and were asked to provide an answer from three choices (``A is better than B'', ``B is better than A'', or ``A and B are about same''). This setting is fairer and easier for participants to evaluate the relative quality of two grouping results than scoring a single result. The order of the grouping results for comparison was randomized. Fig.~\ref{fig:studydemo} illustrates a snapshot of a side-by-side comparison used in one comparison task, where A is manually grouped by a designer and B is automatically recovered from our technique. Thus, we collected 25 comparisons $\times$ 20 participants = 500 responses. It took approximately 30 minutes on average (range from 25 to 40) for each participant to complete all tasks. After they completed the tasks, we asked them to leave comments for the result of their preferences.

\begin{figure}[!t]
  \centering
  \includegraphics[width=\linewidth]{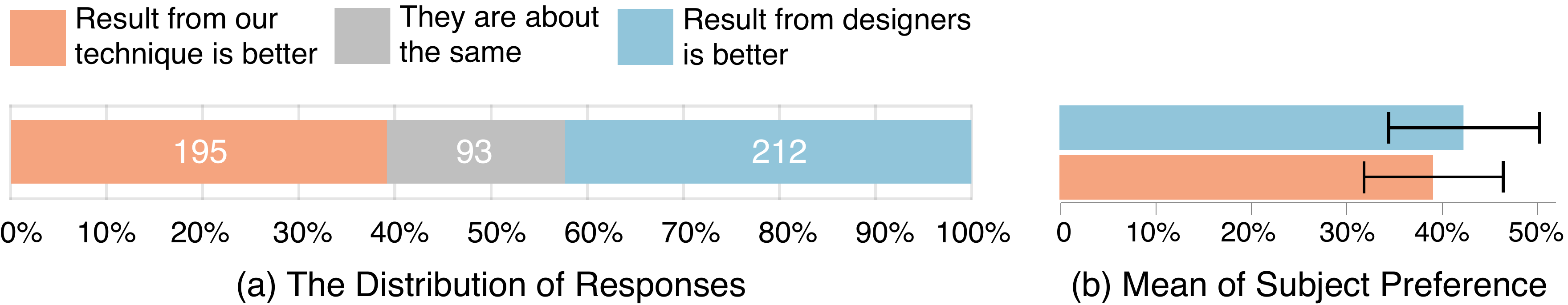}
  \caption{Comparison results. (a) The chart presents the distribution of responses in the comparison experiment; (b) The chart shows mean preference per participant and 95\% confidence intervals.}
\label{fig:comparison}
\end{figure}

\begin{figure*}[!t]
  \centering
  \includegraphics[width=\linewidth]{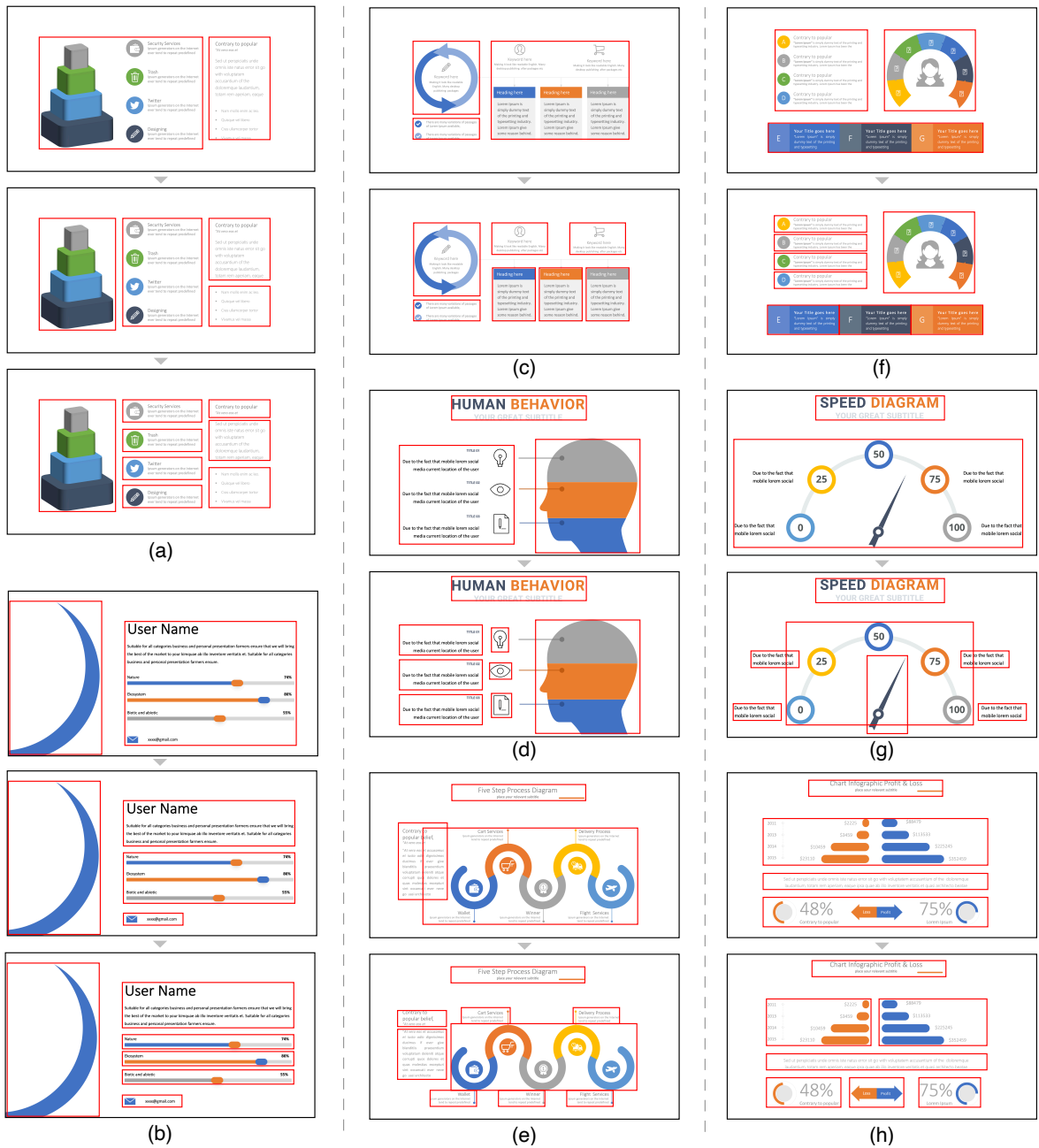}
  \caption{Eight examples of hierarchical groupings recovered by our technique. (a) and (b) have three hierarchies, and the others (c)-(h) have two hierarchies.}
\label{fig:gallery}
\end{figure*}

\textbf{Quantitative Results}
We analyzed the comparison results to investigate the preferences in terms of grouping quality. In general, the preferences for both sides are balanced, as demonstrated in Fig.~\ref{fig:comparison}(a). Specifically, 39\% of the responses (195/500) remarked that the grouping results from our technique are better, and 42.4\% of the responses (212/500) indicated a preference for the results grouped by designers. The rest of the responses (18.6\%, 93/500) indicated that both results are approximately the same. 
Then, we further investigated the preferences of each participant. The mean of preferences per participant and 95\% confidence intervals indicated that the preference for the designer's grouping results is only slightly higher than our technique's (Fig.~\ref{fig:comparison}(b)). According to the Student t-test ($t(19)=0.45, p=0.66$), there was no significant difference between the preferences for both sides. 
On the other hand, our technique can save the extra time of manual grouping, which takes about 30 minutes ($M=30.1, SD=6.69$) for designers to group elements in a ten-page slide in our user study.
As a result, our technique can significantly improve efficiency and produce hierarchical grouping results comparable to human work.

\textbf{Qualitative Results}
During the study, participants were not informed that one in each testing pair was generated by an algorithm. After the study, we asked their opinions about the two results. All participants commented that they did not notice that the grouping was completed by machine. In participants' opinion, most of the grouping results were about the same, but they might select one of them according to their personal preferences. For example, \textbf{P4} commented: ``\textit{Generally, there are no big differences between the two sides. However, when I look at a graphic design, I will think about how to refine the layout. Therefore I would love to choose the proper grouping that may make my design process more convenient.}'' \textbf{P12} said: ``\textit{Most differences are very slight for me to give a choice. I think it is reasonable because designers may have different opinions when grouping the elements}''.
After we told them that one side was from an automatic approach, we asked them to comment on the effectiveness and usefulness of the technique. Eighteen participants appreciated the technique that might significantly improve the design efficiency. Furthermore, some of them offered some suggestions for future works. For example, \textbf{P9} commented that it would be more convenient if each group could be automatically given a proper group name. \textbf{P7} said: ``\textit{I hope the technique can give personalized suggestions by learning from my grouping behaviors. For example, I can demonstrate operations in one design file. Then, I want all my downloaded public design resources to be processed as I demonstrated.}''
Although most participants gave positive feedback, two participants worried that the automatic grouping of elements might limit the creativity of the design to some extent. One of them (\textbf{P10}) said that the grouping structure may affect the designer's strategy to the downstream tasks of layout design.

\subsection{Sample Grouping Results}

The gallery in Fig.~\ref{fig:gallery} demonstrates eight hierarchical grouping results recovered by the automatic approach. (a) and (b) have three hierarchies, and the others (c)-(h) have two hierarchies. Our automatic approach has the ability to group visual elements properly at each level of granularity.
\section{Discussion}

\subsection{Failure Cases and Limitations}

\begin{figure}[!t]
  \centering
  \includegraphics[width=\linewidth]{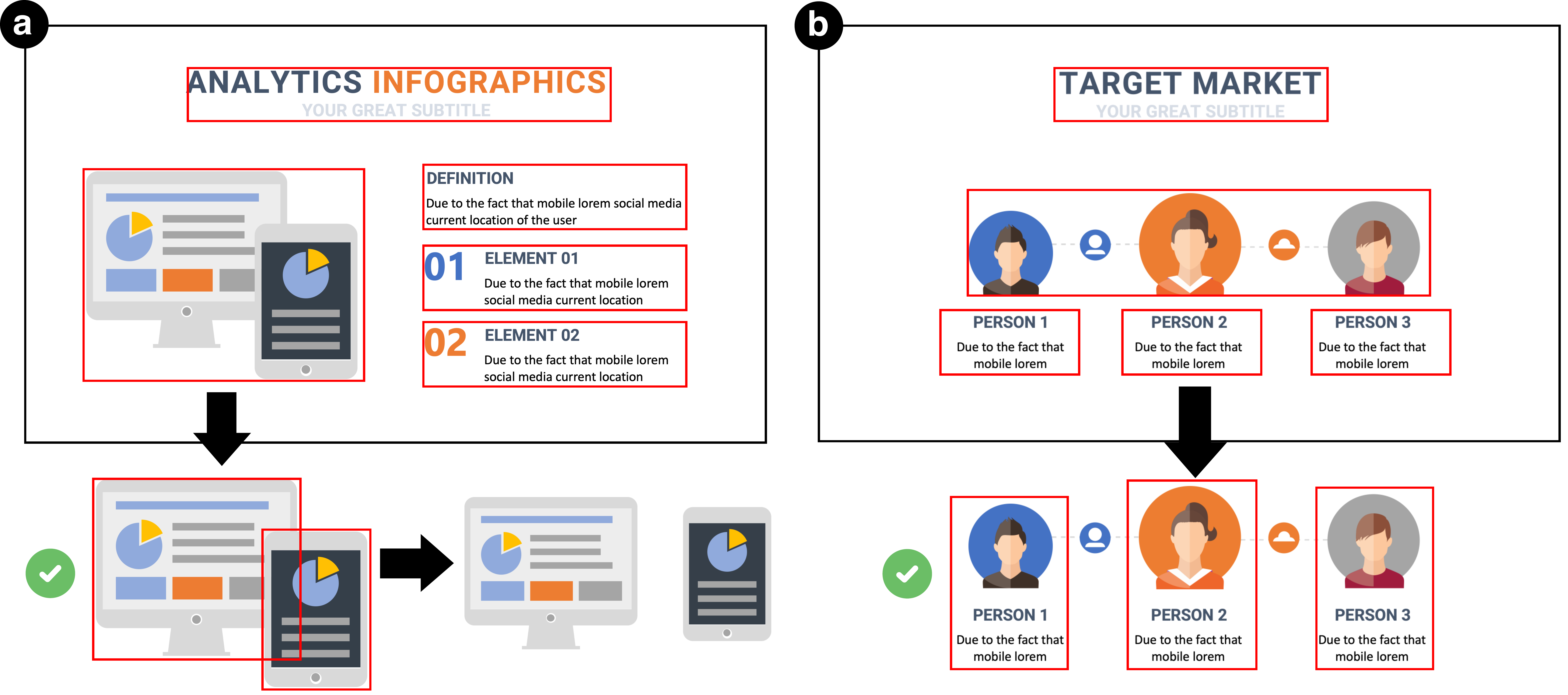}
  \caption{Failure cases: (a) our technique failed to segment two semantic groups of \textit{freeform shape}s;  (b) it was affected by connectors.}
\label{fig:failures}
\end{figure}

We observe some failure cases during the evaluation. We summarize two main causes of the failure. 
One typical failure is that the current approach cannot understand the semantic meaning of a user-defined shape. For example, Fig~\ref{fig:failures}(a) shows a user-defined shape on the left that is composed of multiple \textit{freeform shape}s. Our technique favors grouping them all into a single group at each hierarchical level. However, the designers in our user studies prefer to split it into two parts as a ``computer" and a ``phone" at the lowest level of grouping hierarchy. Although it is easy for human beings, our current technique cannot understand the semantic meaning of these shape compositions. 
Another primary cause of failure is that some complex connectors may affect the grouping accuracy. Simple connectors are lines or arrows to connect visual elements and express their relations. However, the complex one might be composed of multiple elements. For the example in Fig~\ref{fig:failures}(b), the connectors among three head portraits are a composition of a dashed line (\textit{line}), a circle (\textit{preset shape}), and an icon (\textit{freeform shape}). Designers can recognize these connectors and group the head portraits with the corresponding text below. However, our current technique groups all the above visual elements as a whole by mistake. We believe that feeding more training data can mitigate this problem if the model can see more semantic groups.

Our technique also has some limitations in terms of our research scope. 
The first limitation is that our current approach does not involve the content of each visual element, covering text, picture, and shape. Understanding the semantic meaning of the content may assist in understanding the layout structure, which is a promising future work. In addition, we could involve more semantic features, such as role labels (e.g., title, subtitle, and body).
The second limitation is related to user preferences. As we found in the user study, there are multiple ways to group visual elements. Even two professional designers may have different strategies to perform the element grouping task. Our technique can provide the grouping results at different granularities but fails to adjust the strategy according to the user preference. As a participant in the user study mentioned, a promising area for future research might be learning from the user's design behaviors.
Third, this paper does not attempt to build a complete perceptual model of visual grouping. Instead, we propose a learning-based method for hierarchical grouping as a design aid within the domain of information presentations. To understand how designers group the visual elements, more theoretical research on visual perception should be studied further.

\subsection{Usage Scenarios}
\label{sec:usage}

    \textbf{1) Usage of the dataset} 

We see our dataset of information presentations as a useful starting point for exploring layout structures in graphic design. In addition to our grouping research, this dataset is also helpful for researchers in other tasks. For instance, it can help researchers understand the design demographics of information presentations as the collected data are created by various designers. In addition, it can enable a series of studies for design-based machine learning, such as layout clustering, layout classification, layout recommendation, and layout generation. Moreover, this dataset can also serve as a rich design resource to inspire designers when designing layouts.

\textbf{2) Usage of the technique} 

The technique of automatic hierarchical grouping can be applied to the design resources in the wild. Although professional graphic design tools (e.g., PowerPoint and Adobe Illustrator) provide the grouping function for users to organize the grouping structure, the given grouping information in the existing information presentations can be very noisy as designers might only care about the final presentation during the design process. Some unintentional actions or mistakes may also affect the grouping results. Our technique can fix this problem by producing hierarchical groupings for a large batch of information presentations in a short time.

\section{Conclusion}

In this paper, we present an automatic approach for reverse-engineering information presentations. We first collected a large-scale dataset with 23,072 well-designed slides as our research resource. Based on the dataset, we propose a technique to recover hierarchical grouping from layouts of visual elements. Our technique incorporates the advanced deep learning model to understand the relatedness between visual elements. It also includes a bottom-up graph-based algorithm to produce the hierarchical grouping by considering the relatedness and spatial distances. A technical experiment and a user study evaluate the power of the technique, which demonstrates comparable ability with human designers. We believe that future applications can benefit from our contributed dataset and technique.

\bibliographystyle{ACM-Reference-Format}
\bibliography{main}

\end{document}